# Estimation of marine fishing capacity of China

Yi Zheng

*Business School, Shanghai Jian Qiao University, Shanghai 201306，China.*
Email: yzheng@shou.edu.cn

**Abstract**

By the methods of PTP and DEA, the marine fishing capacity in China was studied in quantity. The results showed that: it is better that the input index in the PTP methodology was considered as total power but not as amount of ships in measuring the Chinese marine fishing capacity. The further analyses made it clearer that the excess of marine fishing capacity in China is first the excess of total power to fishing ships. Then, this paper suggested that the amount of ships be reduced 35.2 percent, the gross tonnage be reduced 29.8 percent and the total power be reduced 37.3 percent in China, if the nowadays-marine catch is increased by null on the basis of catch in 1999. In discussion the ideal of supplement peak year was proposed. This paper also discovered that the PTP methodology is suitable to market economics and is better as a longitudinal analysis in time arrays, but the DEA approach is skilled in the transverse comparison of fishing capacity in a same time period. The analyses showed also: the capacity, which has been based on the practical catch, is commonly low estimated. So the fishing capacity reduced in practice is generally larger than these value of calculation.

1. Introduction

China is a fishing country. Its output of the aquatic product has been ahead of the world since 1989. So development of China's fishing industry will make a notable impact on the sustainable fishery in the world. On the other hand, since the late of 70s, in Chinese coastal waters the resources of some major economic species, such as long hair-tail fish, large yellow croaker, small yellow croaker and cutter fish, have fallen or collapsed. The current high output has been occupied by the species with low value or in low position of food chain in chief. Obviously, for protecting the world fishery environment and restoring China's coastal fishing resources, it is necessary to assess China's marine fishing capacity urgently and systematically.

In this paper, by the methods of peak-to-peak ( PTP ) and data envelopment analyses ( DEA ), which are recommend with emphasis by FAO (1988), the authors analyze and quantify the China's marine fishing capacity. These results showed: it is better that the input index in the PTP methodology was considered as total power but not as amount of ships in measuring the Chinese marine fishing capacity. The further analyses made it clearer that the excess of marine fishing capacity in China is first the excess of total power of fishing ships. If the nowadays-marine catch in China is increased by null on the basis of catch in 1999, this paper suggested that the amount of ships be reduced 35.2 percent, the gross tonnage be reduced 29.8 percent and the total power be reduced 37.3 percent. Obviously, these results will be helpful to direct to set up the system of monitoring and controlling fishing capacity in China. Then in discussion the problems in application of PTP methodology were explored, the improving advises were suggested. It is called "supplement peak year". This paper also discovered that the PTP methodology is only suitable to market economics. Comparing the PTP and the DEA method, it is clear that PTP methodology is helpful for a longitudinal analysis in time arrays, but DEA approach is skilled in the transverse comparison of fishing capacity. So in practice application, the PTP methodology should be based upon comparison in the time arrays and the DEA approach should note the transverse comparison in a same time point. The analyses showed also: the capacity, which has been based on the practical catch, is always low estimated. That is, the fishing capacity reduced in practice is generally larger than these values derived by calculation.

2. The outline of concepts and the methods

2.1 The basic concepts

Though much discussion and extensive reference, the technical working group on the management of fishing capacity of FAO agreed with the definition as follows ( FAO, 2000 ):

Fishing capacity is the maximum amount of fish over a period of time (year, season ) that can be produced by a fishing fleet if fully utilized, given the biomass and age structure of the fish stock and the present state of the technology. Fishing capacity is the ability of a vessel or vessels to catch fish.

The measures of fishing capacity might be expressed in term of either input or an output orientation. Input oriented approaches examined the minimum inputs to achieve a given level of output. Output oriented approaches aimed to maximize the potential output given a fixed level of inputs. Measured by output oriented approach, the fishing capacity is called production capacity. Otherwise, Measured by input oriented approach, it is called physical capacity.

The fishing experts emphasize that the production capacity was a "best practice frontier", that is, it reflected the production capacity of the most efficient vessels for the time period analyzed. For this reason, the authors studied from the angle of production capacity to estimate Chinese fishing capacity in this paper.

Capacity utilization is the ratio of actual output (Y) to some measure of potential output (Y*) given a firm's short-run stock of capital and perhaps other fixed inputs in the short run, i.e. Y/Y*(Nelson. 1989).

2.2 The peak-to-peak ( PTP ) methodology

The PTP methodology is a direct and simple measurement of the institutionalized or observed response by the industry to changes in demand ( Ballard and Roberts, 1977 ). It is based on a "trend through peaks" approach that is thought to reflect maximum attainable output given the stocks of capital and fish. Its demand on data is the lowest in all methods of the assessment of fishing capacity and capacity utilization. The specification used in the empirical analysis is the equation:

$$Y_t/V_t = AT_t \qquad \text{--------(1)}$$

Where $Y_t$ represents total catch or output which can be produced in the current time period. $V_t$ is a single production unit, which have been combined by the labor and capital. A is a constant coefficient. $T_t$ is an adjusting technology trend; it is estimated as follow:

$$T_t = T_{t-m} + \frac{Y_{t+n}/V_{t+n} - Y_{t-m}/V_{t-m}}{n+m} * m \qquad \text{--------(2)}$$

Relative to a particular year t, the values of m and n correspond to the length of time from the previous and following peak years.

### 2.3 The Data Envelopment Analysis (DEA)

Of the various approaches, the DEA approach perhaps is the easiest and offers the most promising and flexible method to determined capacity and capacity utilization. DEA has been widely applied to problems in which answers about optimum input levels, their characteristics, and output levels were desired ( FAO fisheries department, 1999).

Assume there is data on K inputs and M outputs on each of N firms or DMU's (decision making units) as they tend to be called in the DEA literature. For the i-th DMU the vectors $x_i$ and $y_i$ represent these respectively. The K×N input matrix, X, and the M×N output matrix, Y, represent the data of all N DMU's. Given the constant returns to scale (CRS) assumption, the input-oriented DEA model can be derived as follows:

$$\begin{aligned}
&\text{Min}_{\theta,\lambda}\ \theta \\
&\text{s.t.}\ -y_i + Y\lambda \geqslant 0 \\
&\quad\ \ \theta x_i - X\lambda \geqslant 0 \\
&\quad\ \ \lambda \geqslant 0
\end{aligned} \qquad \text{---------(3)}$$

Where θ is a scalar and λ is a N×1 vector of constants. For i=1, 2…N, the value of θ obtained from the solution of the i-th linear programming problem will be the efficient score for the i-th DMU. It will satisfy $0 \leq \theta \leq 1$, with a value of 1 indicating a point on the frontier and hence a technically efficient DMU, according to the Farell (1957).

The output-oriented DEA model is similar to its input-oriented counterparts, that is:

$$\begin{aligned}
&\text{Max}_{\varphi,\lambda}\ \Phi \\
&\text{s.t.}\ -\Phi y_i + Y\lambda \geqslant 0 \\
&\quad\ \ x_i - X\lambda \geqslant 0 \\
&\quad\ \ \lambda \geqslant 0
\end{aligned} \qquad \text{---------(4)}$$

Where $1 \leqslant \Phi \leqslant +\infty$, and Φ is the proportional increase in outputs that could be achieved by the i-th DMU, with input quantity held constant. Note that 1/Φ defines a TE (technical efficiency) score which varies between zero and one. In the research of capacity TE is also the capacity utilization.

## 3. Analysis of marine fishing capacity in China

### 3.1 Estimation of Chinese marine fishing capacity by the PTP methodology

For studying PTP's suitability in China, this paper chose the fisheries of the city Zhousan and whole country to represent small and large fishing area. According to the data of their marine catch, number and total power of the fishing vessels (see table 1 and table 2), the marine fishing capacity in Zhousan and whole country has been calculated by the PTP methodology ( formula (1) and (2) ). The calculated values and the practice output were compared in figure 1 and figure 2. Where $Y'_t$ represents practical catch, $Y'_{1t}$ is capacity output calculated by the number of vessels as an input index, $Y'_{2t}$ is capacity output calculated by the total power (kW) as input, $Y'_{3t}$ is capacity output calculated by the gross tonnage.

In figure 1 it can be found that the capacity output $Y'_{2t}$ is almost similar to the capacity output $Y'_{3t}$. So it is enough to choose one as input index between total power and gross tonnage. If we consider the data of total power is more available and more correct than the data of gross tonnage in China, it is better to choose total power as an input index. On the other hand, figure 1 and figure 2 are all showed: the value of $Y'_{1t}$ with the input index of number of vessels was not always reasonable, because it was less than the practice catch $Y'_t$ in some year. It contradicts the definition of the fishing capacity, which ought to be the maximum potential output.

Table 1 Data about marine fishing in Zhousan

| Year | Practice output (ten thousand t) ($Y_t$) | Marine engine vessels | | |
|---|---|---|---|---|
| | | ten thousand vessels $V_{1t}$ | ten thousand kW $V_{2t}$ | ten thousand gross tonnage $V_{3t}$ |
| 1995 | 88.81 | 1.22 | 43.41 | 147.40 |
| 1996 | 96.06 | 1.20 | 45.02 | 152.85 |
| 1997 | 105.90 | 1.19 | 45.30 | 155.05 |
| 1998 | 123.35 | 1.15 | 49.25 | 168.19 |
| 1999 | 123.07 | 1.11 | 50.81 | 174.31 |

Source: Fishery of Zhousan.

Table 2 Data about marine fishing in China

| year | The practical output of marine fishing in coastal waters (ten thousand t) ($Y_t$) | Marine engine vessels in coastal waters | |
|---|---|---|---|
| | | The number of vessels $V_{1t}$ | kW $V_{2t}$ |
| 1993 | 851.75 | 252,126 | 8,106,593 |
| 1994 | 994.44 | 259,297 | 8,394,107 |
| 1995 | 1139.75 | 273,978 | 9,800,739 |
| 1996 | 1245.64 | 280,352 | 10,755,063 |
| 1997 | 1385.38 | 282,504 | 11,218,769 |
| 1998 | 1496.68 | 283,218 | 11,801,492 |
| 1999 | 1497.62 | 279,994 | 12,180,709 |

Source: Statistics of Chinese Fishery

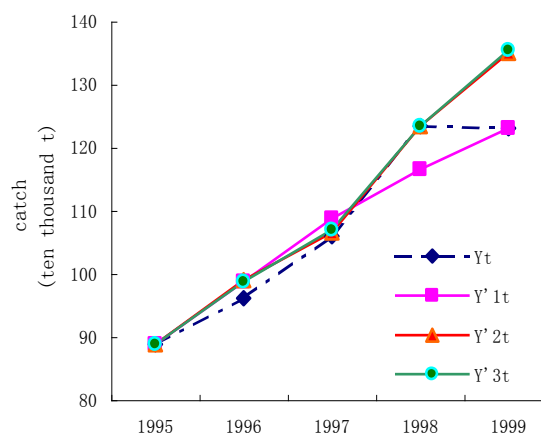

Fig.1 The practical catch and the capacity output in Zhousan

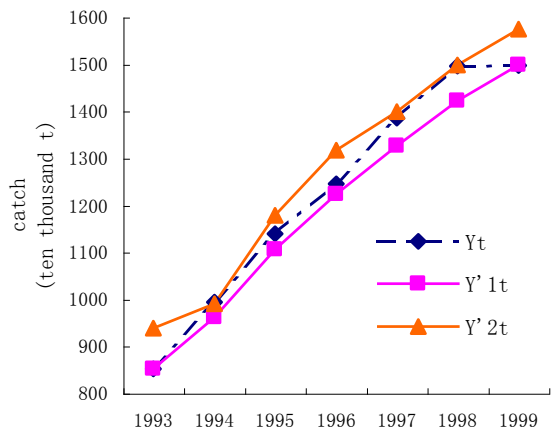

Fig.2 The practical catch and the capacity output in China

Base these analyses the authors recommend: The total power is better as an input index for measurement of marine fishing capacity in China by PTP methodology, although the number of vessels has often been applied in many relative literatures.

3.2 Estimation of Chinese fishing capacity in 1999 by the DEA approach

According to the materials of Statistics of Chinese Fishery, making the Chinese coastal provinces as MDUs, the authors assessed the fishing capacity of Chinese marine fleets on the assumption of CRS by the output-oriented and input-oriented DEA models ( formula (3) and (4) ). The detailed calculation results are listed in table 3.

The value in table 3 derived by output-oriented DEA showed: Chinese practice marine catch in 1999 is 1,497.62 ten thousand tons, but the capacity output for current input is 2,124.05 ten thousand tons. The capacity utilization is 69.1%. It is obvious that the capacity didn't utilize enough. The major reason is that the current input in Chinese marine fishery is in large excess of the endurance of the resources. That is, the excess of capacity in China is positive.

Chinese government demanded its marine fishing for null-increase to catch in table in 1999. This is a directing policy for controlling the country's marine fishing capacity. So to Chinese practical marine catch in 1999, this paper assessed the required minimum input for its all coastal provinces and whole country (see the value by input-oriented DEA in table 3). From these results, it is easy to know how much current input is excessive and how much should be reduced in the case of null-increase.

In the concrete, if the nowadays-marine catch is increased by null on the basis of catch in 1999, Chinese number of vessels should be reduced from the current 280,550 to 181,802. The rate of reducing is 35.2 percent. The gross tonnage should be reduced from the current 5,722,277 to 4,017,198. The rate of reducing is 29.8 percent. The total power of vessels should be reduced from the current 12,419,238 kW to 7,788,893 kW. The rate of reducing is 37.3 percent. For every coastal province the minimum input for getting current catch could see the value by the input-oriented DEA in table 3.

Table 3  Calculation of marine fishing capacity in China in 1999

| | Practical data of marine fishing | | | | The value by output-oriented DEA | | The value by input-oriented DEA | | |
|---|---|---|---|---|---|---|---|---|---|
| | Catch (ten thousand) | Number of vessels | Gross tonnage | kW | Capacity utiliza-tion | Capacity output (ten thousand) | Number of vessels | Gross tonnage | kW |
| Tianjin | 3.41 | 1082 | 33245 | 57274 | 35.4% | 9.61 | 383 | 10455 | 20295 |
| Hebei | 32.83 | 9008 | 169438 | 338483 | 46.6% | 70.51 | 4194 | 78892 | 152802 |
| Liaoning | 157.69 | 32679 | 409602 | 891422 | 81.2% | 194.13 | 21194 | 332727 | 643529 |
| Shanghai | 12.20 | 863 | 102321 | 173271 | 77.1% | 15.82 | 666 | 68604 | 133647 |
| Jiangsu | 68.32 | 19852 | 379421 | 743745 | 43.8% | 156.15 | 8685 | 166000 | 321552 |
| Zhejiang | 331.24 | 39851 | 1949320 | 3684936 | 72.9% | 454.65 | 29034 | 1380198 | 2684710 |
| Fujian | 206.65 | 52718 | 640024 | 1724995 | 68.1% | 303.33 | 27774 | 436020 | 843310 |
| Shandong | 332.52 | 44692 | 701610 | 1356990 | 100.0% | 332.52 | 44692 | 701610 | 1356990 |
| Guangdong | 194.52 | 53497 | 816838 | 2241062 | 50.2% | 387.13 | 26144 | 410426 | 793809 |
| Guangxi | 88.86 | 11753 | 204561 | 493090 | 98.8% | 89.94 | 11611 | 202085 | 391181 |
| Hainan | 51.10 | 13999 | 194039 | 475441 | 55.6% | 91.96 | 6868 | 107822 | 208540 |
| Aquatic productions Co. | 18.30 | 556 | 122358 | 238529 | 100.0% | 18.30 | 556 | 122358 | 238529 |
| Whole China | 1497.62 | 280550 | 5722777 | 12419238 | 69.1% | 2124.05 | 181802 | 4017198 | 7788893 |

After the above analyses, it is clear that the excess of marine fishing capacity in China is first the excess of total power. This is say; the increase of Chinese marine catch has depended in chief on the increase of total power. So this paper got the conclusion: for reducing the fishing capacity of Chinese marine fleets, it is first to control total power of the fleets. Then the number of vessels and gross tonnage should be controlled too.

The further analyses showed: the capacity utilization is different in different fishing area. So the reducing of fishing capacity should not be same too. In the waters where the capacity utilization is lower, the extent of reducing is greater. But the fisheries, for example the distant-water fisheries, whose capacity utilization is high, should be protected and farther developed.

4. Discussion

4.1 The suitability of PTP methodology in China marine fisheries
4.1.1 Choosing of peaks in PTP

It is important for the calculating results to choose appropriate PTP's peaks, which is the year when the rate of productivity of unit input is maximal nearby. But in application, the authors found the peak's definition is not all right. As an example, a set of practical catch and input data were constructed in table 4. Observed the table, it is maximal in No.5 year to the practice catch of unit input $Y_t/V_t$. So No.5 year should be a peak year by the PTP methodology. Calculated based on this, the technology trends $T'_t$ (i.e. capacity output of unit input) is less than the corresponding $Y_t/V_{1t}$ in No.4 year. The capacity output 1 ($Y'_t$) will also less than the practical catch ($Y_t$). This contradicted the definition of the capacity. For settlement of this issue, this paper proposed the supplementary peak years to complete the PTP methodology.

Tab.4 Comparison between the fishing capacity calculated on different peeks

| year | practical catch (Yt) (ten thousand) | input ($V_t$)(ten thousand) | $Y_t/V_t$ | $T'_t$ | capacity output 1 (Y't)(ten thousand) | the changing rate (a) | $T''_t$ | capacity output 2 (Y"t)(ten thousand) |
|---|---|---|---|---|---|---|---|---|
| No.1 | 80.00 | 2.00 | 40.00 | 40.00 | 80.00 | | 40.00 | 80.00 |
| No.2 | 90.00 | 2.00 | 45.00 | 45.63 | 91.25 | 5.00 | 46.67 | 93.33 |
| No.3 | 100.00 | 2.00 | 50.00 | 51.25 | 102.50 | 5.00 | 53.33 | 106.67 |
| No.4 | 120.00 | 2.00 | 60.00 | 56.88 | 113.75 | 6.67 | 60.00 | 120.00 |
| No.5 | 125.00 | 2.00 | 62.50 | 62.50 | 125.00 | 5.63 | 66.67 | 133.33 |
| No.6 | 120.00 | 2.00 | 60.00 | 68.13 | 136.25 | 4.00 | 73.33 | 146.67 |

In addition, for linear relation to reflect the changing of the technology trend, the farther between two peaks, the less reliable the results will be. So the supplement of peak years is helpful to increase the reliability of the PTP methodology, because it may shorten the distance between the peak years.

The method of supplement of peak years made it essential to supplement these years as new peak years, if the changing rate of these peak years to previous peak for the output of the unit input, i.e.

$$a = \frac{Y_{t+n}/V_{t+n} - Y_t/V_t}{n} \quad \text{---------- (5)}$$

is maximal nearby. In equation (5), t represents the year of previous peak, t+n represent the year, which is n year apart from the previous peak year.

Taking the example of table 4, the changing rate (a) is maximal in No.4 year (see table 4), so the year is supplemented as new peak year. Based this the technology trends ($T''_t$) is more than or equal to the corresponding practical catch of unit input ($Y_t/V_{1t}$) in every year. As a result, the capacity output 2 ($Y''_t$) is also more than or equal to the corresponding practical catch ($Y_t$). Hence the preceding problem can be solved completely.

### 4.1.2 The suitable scope of the PTP methodology

When the analyses has been done to Chinese 50-60s marine fishing fleets by PTP methodology, there is a problem, i.e. the technology trends $T_t$ are reduced as showed in figure 3. Obviously it didn't conform to reality. Because the technology trends, which are decided by the level of the same year's technology, shouldn't at least become less than the former.

For resolving the problem, the authors reviewed the limit conditions of PTP methodology. It has been found: To the base of the PTP methodology, which is the Cobb-Douglas's' production and the assumptions that labor and capital are fixed proportion over time, it is more reasonable in the presupposition of market economics. This can be further proved from the calculation in this paper too. In 50-60s China is in an era of plan economics, the results (figure 3) showed the PTP methodology was not suitable. In 90s the market economics was hold main position in Chinese marine fishing, the results calculated by PTP (see figure 1and figure 2) were comparatively ideal.

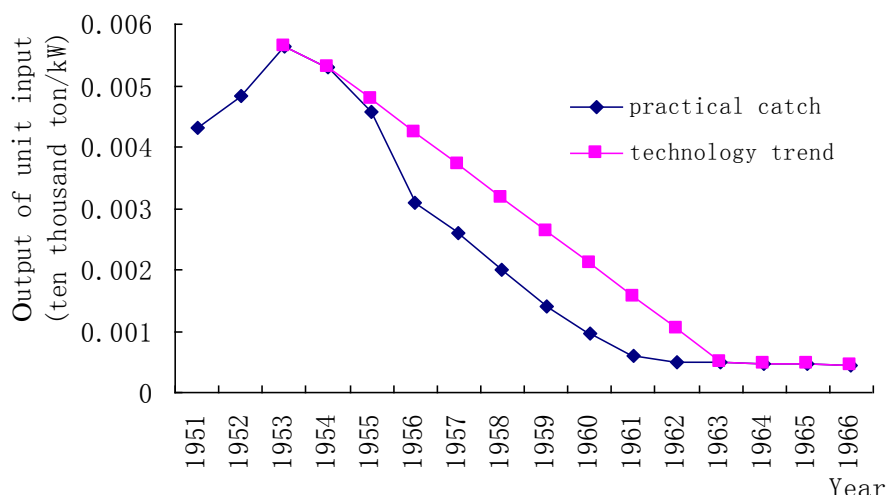

Fig.3 Output of unit input in Chinese marine fleets from 1951 to 1966

Although there may be other methods applied to research in this area in the future (Zheng et al., 2022), the above analyses confirmed that PTP methodology is more suitable to the marine fishing in the perfect market economics.

### 4.2 The comparison between the two methods of measuring capacity

Analyzed the application of the two methods, this paper found: the PTP methodology can reflect the variations of the fishing technology to a certain extent in a time array. In the estimation of the capacity to a fishing unit, which has only a kind of input, the results derived by PTP in a time array are more reasonable than the DEA's. But the PTP methodology is only used in the analyses of one kind of input of a fishing unit and it must satisfy the assumption of CRS. So it is quite limited. On the contrast, the DEA approach can be applied not only to a lot of kind of inputs but also to many fishing units. In addition, it can be calculated not only in CRS but also in VRS (variable returns to scale). Further, the DEA approach may put together the information of the many fishing units in same time period, so it is more reasonable and more correct than the PTP methodology to estimate the fishing capacity in a time period. The DEA approach can also make sensitivity analyses for the factors, which has affected the fishing capacity. It is impossible to the PTP methodology.

In brief, PTP methodology is a longitudinal analysis in time arrays, but DEA approach is skilled in the transverse comparison of fishing capacity. Based these analyses, the PTP methodology may be used to predict the capacity in coming years and the DEA approach is more ideal to estimate the capacity of a certain year.

4.3  Paying attention of the appliance for results of the calculation in the practice

Calculated by the DEA approach, this paper suggested that in China the amount of ships be reduced 35.2 percent, the gross tonnage be reduced 29.8 percent and the total power be reduced 37.3 percent, if the nowadays-marine catch is increased by null on the basis of catch in 1999. Because the resources have declined at present, the capacity, which has been based on the practical catch, is always low estimated. In addition, considering reduction of the capacity and restoration of the resource, the fishing efficiency may be farther raised. So the fishing capacity reduced in practice is generally larger than these above values are. It is necessary to adjust the reducing rate continuously on the basis of new circumstances.

Secondly, from the figure 2 and the table 3, in 1999 the capacity output derived by the PTP methodology is about 1500 ten thousand ton and by the DEA approach is about 2100 ten thousand ton. The distinction between them also showed: the results derived by PTP methodology only have the meaning to compare in time arrays. But the DEA approach can further describe the fishing capacity in a time period. For this season, to analyze the practical problems, it should be based upon comparison in a time array for applying of the PTP methodology and should note the transverse comparison in a same time period for applying of the DEA approach. So it is very important to pay attention to the difference for applying the two methods appropriately in practice.

References
FAO. 1998. Report of the technical working group on the management of fishing capacity. FAO Fisheries Report No. 586, 13.
FAO fisheries department. 1999. FAO fisheries technical paper 386. 106.
FAO. 2000. Report of the technical working group on the management of fishing capacity. FAO Fisheries Report No. 615, 32-51.
Nelson, R. 1989. On the measurement of capacity utilization. Journal of Industrial Economics, Vol. XXXVII, No.3.
Ballard K and J. 1977. Roberts. Empirical stimulation of the capacity utilization rates of fishing vessels in 10 major Pacific coast fisheries. Washington, D, C: National Marine Fisheries Service.
Farell, M.J. 1957. The measurement of productive efficiency. Journal of the royal statistical society. A CXX, Part3, 253-290.
Zheng, J., Makar, M. (2022).Causally motivated multi-shortcut identification & removal. Advances in Neural Information Processing Systems 35,12800-12812.